\documentclass[journal]{IEEEtran}

\usepackage{listings}

\usepackage{graphicx}
\usepackage{multirow} 
\usepackage{array}
\usepackage{overpic}

\usepackage{hyperref}
\hypersetup{hypertex=true,
            colorlinks=true,
            linkcolor=blue,
            anchorcolor=blue,
            citecolor=blue}
\usepackage{tcolorbox}

\usepackage{amsmath} 
\usepackage{amssymb}
\usepackage[T1]{fontenc}
\usepackage{booktabs} 
\usepackage{multirow}
\usepackage{threeparttable}

\usepackage{amsfonts,amssymb}
\usepackage{pythonhighlight}
\usepackage[utf8]{inputenc}
\usepackage{fancyhdr}

\begin{document}

% >>:-----------------------------------------
% >>:-----------------------------------------
\title{ecVoice: Audio Text Extraction and Optimization of Video Based on Idioms Similarity Replacement}
% >>:-----------------------------------------

\author{Jinwei~Lin,~\IEEEmembership{Member,~IEEE, ACM}
        
\thanks{Jinwei Lin is come from Monash University,} 
\thanks{ORCID of Jinwei Lin:0000 0003 0558 6699,} 
\thanks{Manuscript written in 2018.}} 

% Paper Title
\markboth{Journal or Conference of \LaTeX\, No.~1, May~2023}
{Shell \MakeLowercase{\textit{et al.}}: Simple  Arrow Area Architecture Template}
\maketitle

% <<:-----------------------------------------

% >>:-----------------------------------------
\begin{abstract}
The Text Extraction of the Audio from the Video plays an important role in multimedia editing and processing. As a popular open source toolkit, Whisper performs fast in human voice recognition. However, the recognition performance is dependent on the computing resource, which makes the low computing memory running Whisper become difficult. Our paper presents an available solution to extract the human voice from the video and gain the high quality text generation from the voice. The generated voice can be used in video language translation and translated voice simulation. To improve the extraction and transform quality of human voice, we present ecVoice, a method using the idioms similarity computation and analysis to improve the quality of audio text extraction. Relative experiments are held to verify that the ecVoice can improve the idiom grammar correction rate to 90\% on average. The method is simple but fast which means this method will cause less bad influence of consuming computing resources when improving the voice recognition rate. Our method and solution can significantly enhance the Whisper recognition with low computing memory.
\end{abstract}
% <<:-----------------------------------------

% <<:-----------------------------------------

% >>:===========================================
\section{Introduction}
Extracting and gaining a high quality audio text from the video is a significant application and implementation in multimedia. The text extracted from the video can be used in multiple fields of multimedia applications or research implementations, such as translated language of video playing, audio editing and internationalisation of game video design. With the technologies of super Artificial Intelligence (AI), AI assistants have more and more important roles in multimedia and computer games design. One important process or research problem of these applications is transforming the voice to text, which is also one of the most important points of our research. 

However, there are still some difficulties or challenges in transforming the voice to text, for example, improving the recognition accuracy and processing speed.Usually, methods like Deep Neural Network (DNN) are used to address this issue. But the recognition accuracy may of course have a large fluctuation influence which is related to the parameter size of the neural network model. For some machines or hardware with limited gpu performance or memory, the network parameter size of the model is one of the main influence factors. For the limitation of computing hardwares, one feasible method to improve the recognition performance is using the software solutions, like improving the algorithms design and architecture. In this paper, these software methods can be summarised into three categories, the first is pre-processing methods, the second is the during-processing, the third is post-processing methods. We will focus more on the post-processing method in research of the paper.

In the paper, we will present an available solution to split the audio from the video and extract the accurate text from the audio with an efficient and prolongable algorithm to improve the performance and accuracy of audio text extraction based on idioms similarity replacement. The relative experiment code is open source in GitHub:\href{https://github.com/JYLinOK/ecVoice}{https://github.com/JYLinOK/ecVoice}.

% <<:===========================================

% >>:===========================================
\section{Literature Review}

% >>:-----------------------------------------
\subsection{Extract Text from Voice}
Extracting the text from voice files is related to the research field of voice recognition. The technologies of voice recognition have been used in various fields in AI, such as using speech recognition to recognize the emotion of the speaker \cite{zhang2019spontaneous}, the end-to-end automatic speech recognition \cite{li2022recent}, as well as using different methods to improve the speech recognition accuracy in normal speech recognition filed \cite{shi2022robust}. The efficiency of speech recognition is closely related to the volume of the given information from the voice and how the recognition performs a high quality recognition in limited conditions \cite{fendji2022automatic}. The accuracy of speech recognition can be improved by improving the process algorithms \cite{kumar2022comprehensive}, or using special processes on voice data \cite{green2022speech}. Some researchers try to improve the recognition accuracy by  focusing on the text processing after the post-processing of speech recognition \cite{atmaja2022survey}. 

Our research is focused on extracting the text from the voice in the wild that includes human speech, and improving the speech text recognition by post-process of the recognized text.
% <<:-----------------------------------------

% >>:-----------------------------------------
\subsection{Voice in Multimedia Editing}
In this research, we focus on improving the efficiency of the voice processing in multimedia editing. Voice processing plays an important role as visual information processing in modern multimedia editing \cite{cceken2022multimedia}. Voice processing is also an important part in video editing, and plays an important role in video text translation \cite{diaz2019technological} and game design \cite{a2021ethics}. The audio from the video and then translate the speech voice to text for further multimedia processing \cite{rajeev2023insightful}, which is considered as a significant application in voice recognition. And the research of the paper is focused on how to improve the accuracy of speech recognition from the audio that was extracted from the video, by presenting an algorithm on the post-processing of voice recognition. This research work will be applied and accessed in our further research on automatic intelligent multimedia editing and NeRF AI game engine design in the future. 
% <<:-----------------------------------------

% >>:-----------------------------------------
\subsection{Text Grammar Correction}
After translating the voice to text from the audio, two factors will influence the currency of the speech recognition. One is the recognition success rate of the speech recognition model, which can only be improved by updating and optimising the design or training strategy of the recognition model \cite{he2019streaming}. The other is the pre-processing \cite{tu2019speech} or post-processing and optimization of the translated text from the recognition model. For the post-processing of the translated text, the text syntax correction process in the field of natural language \cite{solyman2021synthetic} is to directly improve the textual expression quality of the translated text to improve the accuracy of the voice recognition.

One of the most important keys of our method to improve the  quality of the audio text extracted from the video is to improve the efficiency of text grammar correction by correcting the wrong expression of the idioms with idioms similarity replacement. Our method is simple but fast, which makes it can be used in the optimising and correcting translated text extracted from the speech audio with a high speed requirement.
% <<:-----------------------------------------
% <<:===========================================

% >>:===========================================
\section{Methodology Analysis}

% >>:-----------------------------------------
\subsection{Running Architecture}
\label{cm1}
\begin{figure}[t]
	\centering
	\includegraphics[width=0.8\columnwidth]{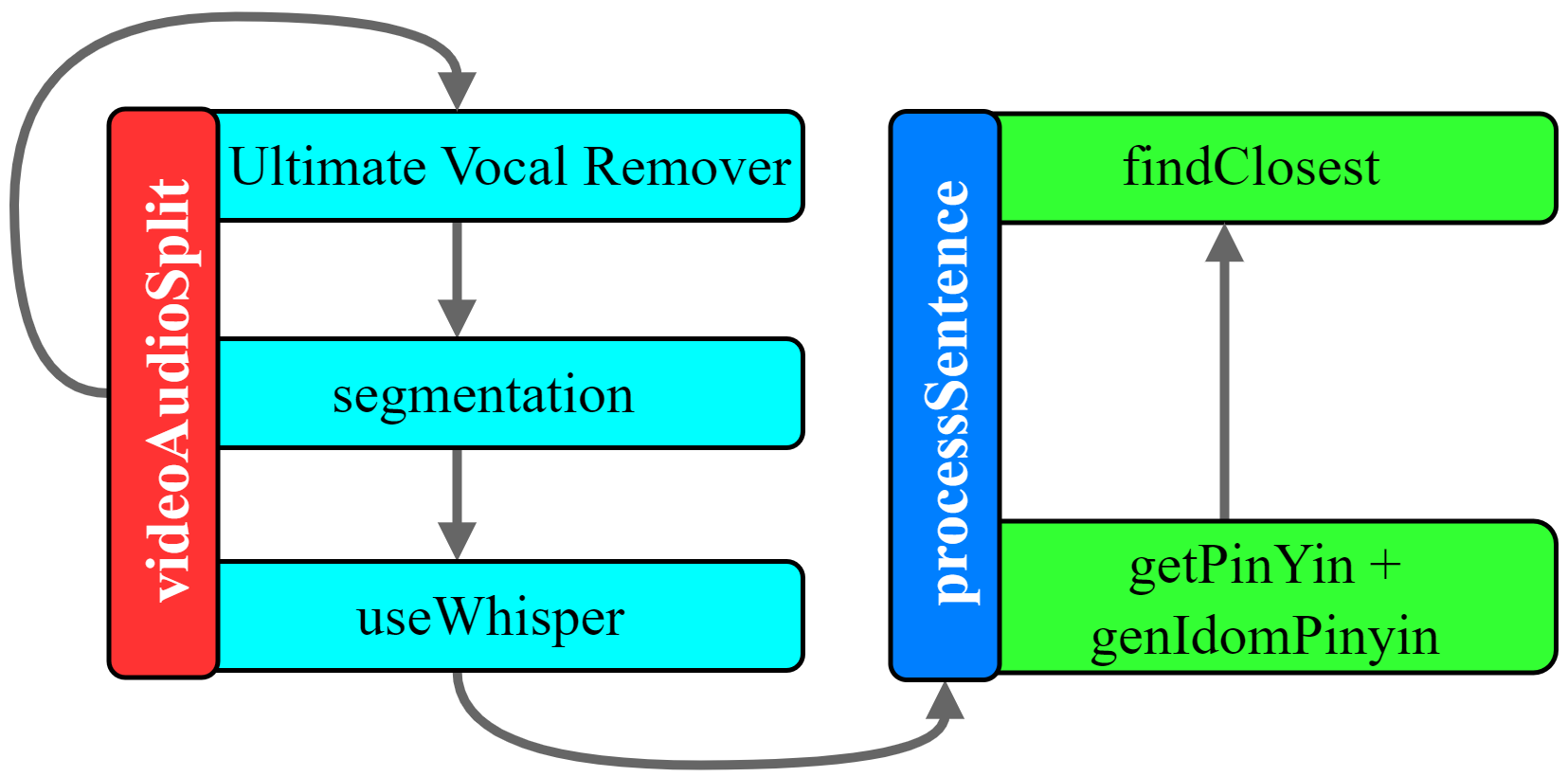}
	\caption{Running and design architecture of this whole research.}
	\label{fig1}
\end{figure}

Optimising textual expression of the extracted text that is translated from the audio of the video is the main process and algorithm analysis of this research. Beside this, splitting the audio data from a specific video, and making the pre-processing of the voice data, are also an important part of the running architecture of the research. As shown in Figure \ref{fig1}, the running architecture of this research includes 7 key components: splitting audio from video, splitting speech from audio, split the main speech, speech text recognition, creating the pinyin analysis function, generating the pinyin dictionary of idioms data-set, finding the closest idioms algorithm. Following are the detailed analysis of the components mentioned above.

As shown in Figure \ref{fig1}, first, process splitting audio from video, is related to the item $videoAudioSplit$. The target of  the process is extracting the audio files from the specific video file. Second, process splitting speech from audio, is related to the item $Ultimate Vocal Remover$. Because the toolkit Ultimate Vocal Remover will be used to gain the pure human speech data with a high quality from the audio files. Third, the process of splitting the main speech is related to the item $segmentation$, which means making the segmentation for the speech file to get the main useful segment data. Fourthly, the processing of speech text recognition is related to the item $useWhisper$, which means using the robust toolkit whisper to generate the accurate translated text of the speech file. Fifthly, the process of analysing pinyin and generating the pinyin dictionary is related to the item $getPinYin+genIdomPinyin$, which means creating function to get the basic information of the segments of words, and generating a pinyin dictionary with a special data format. Last, the process of finding the closest idioms algorithm is related to the item $findClosest$, which means analysing the expression of a segment of words to check if it is a normal idiom with the correct writing format or not, and get the most closest right idiom expression if the words segment is wrong.

% <<:-----------------------------------------

% >>:-----------------------------------------
\subsection{Splitting Audio From Video}
\label{cm2}
\begin{figure}[t]
	\centering
	\includegraphics[width=0.8\columnwidth]{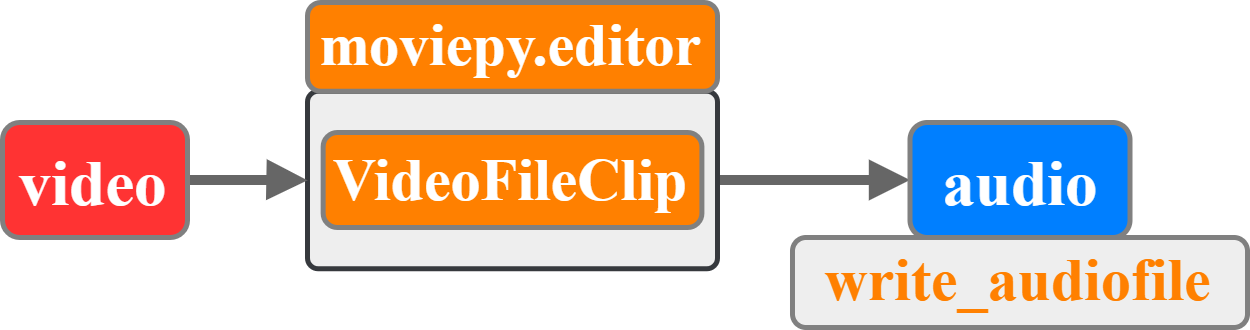}
	\caption{Using Moviepy to splitting the audio data from video.}
	\label{fig2}
\end{figure}

If we want to implement the optimization for the text expression that is extracted from the audio file, the primary process is getting the audio files. Due to focusing on the target of  multimedia processing, the audio files used in this research are all split from the video file, which is corresponding to the item $videoAudioSplit$ in \nameref{cm1}. As shown in Figure \ref{fig2},  we utilise the Python library $moviepy$ to design a processing algorithm to get the split audio file from one specific video file. Using the function $mp.VideoFileClip$ to generate a video clip of an input video with the mp4 format, followed by using the function $write\_audiofile$ to resave the audio data as a wav format file. Thus  getting the audio file of one specific video file. The format of the extracted audio file can be custom defined in the processing code.

% <<:-----------------------------------------

% >>:-----------------------------------------
\subsection{Audio Segmentation}
\label{cm3}
\begin{figure}[t]
	\centering
	\includegraphics[width=1.0\columnwidth]{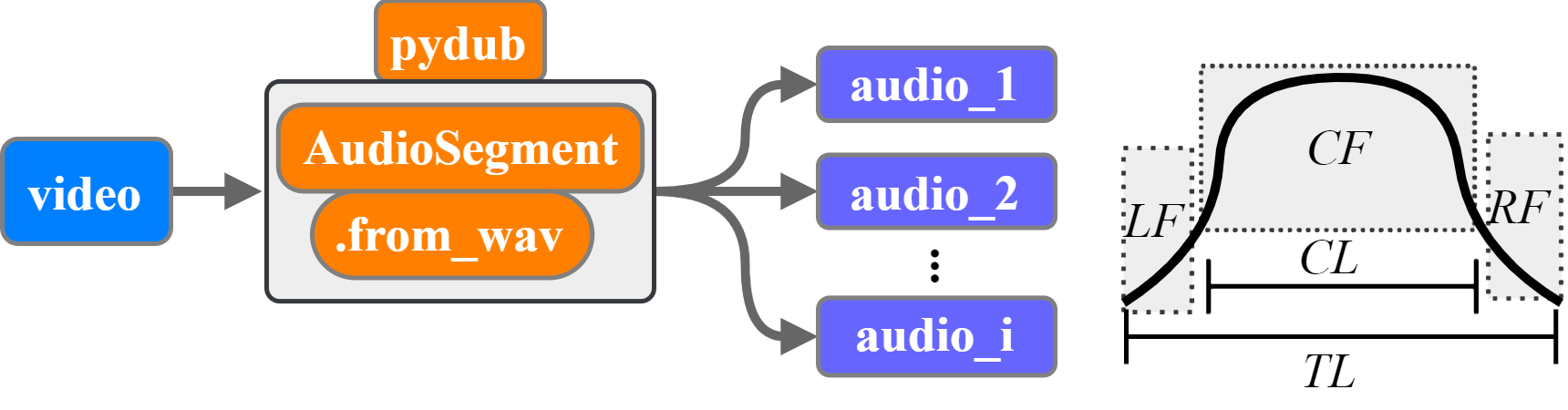}
	\caption{Using pydub to make the segmentation of the extracted audio.}
	\label{fig3}
\end{figure}

After gaining the extracted audio file from the video file, further processing related to the voice optimization can be implemented. In our research, to ensure the high quality of the  audio data before it is processed in future procedures, like splitting the human speech voice and the background audio made by other objects but not human voice, we make the process of cutting off the two ends audio segments of the previous audio that is extracted from the video. This process and design is corresponding to the item $segmentation$ in \nameref{cm1}. As shown in Figure \ref{fig3}, we use the $pydub$ library to make the segmentation of the audio. Detailedly, Using the function $AudioSegment.from\_wav$ to transform the audio file, for example in $wav$ format, subsequently, using the index of list to make the segmentation of the audio file. During this processing, the audio file is presented as one kind of editable data with the list format. After the segmentation edit of the audio, calling the function $export$ of the split audio. Eventually, the previous audio will be split into a specific number of segments.

As shown in Figure \ref{fig3}, assume the total length of audio is represented as STLS, when the audio is generated or recorded, usually the two ends of the audio are sick in recording the information, which will cause some mistakes in information representation. Therefore in this research,  processing algorithm will delete the two ends of the audio to reduce the related interference, and use the middle segment of the audio as the processed audio. As shown in Figure \ref{fig3}, the $LF$ represents the left end of audio length, the $RL$ represents the right end of audio length, the $CL$ represents the central of audio length. 

% <<:-----------------------------------------

% >>:-----------------------------------------
\subsection{Audio to Text}
\label{cm4}
After the segmentation of audio, the next process is translating the speech audio to text, which means the process needs the speech recognition to get the text of the human speech in the audio. To ensure the quality of this process of speech recognition, we use the toolkit $Ultimate Vocal Remover$ to split the human speech from the audio. Except for human speech audio, the background music or voice of the previous audio will be saved for future works, like video or audio speech language translation and multimedia application editing. 

\begin{figure}[t]
	\centering
	\includegraphics[width=1.0\columnwidth]{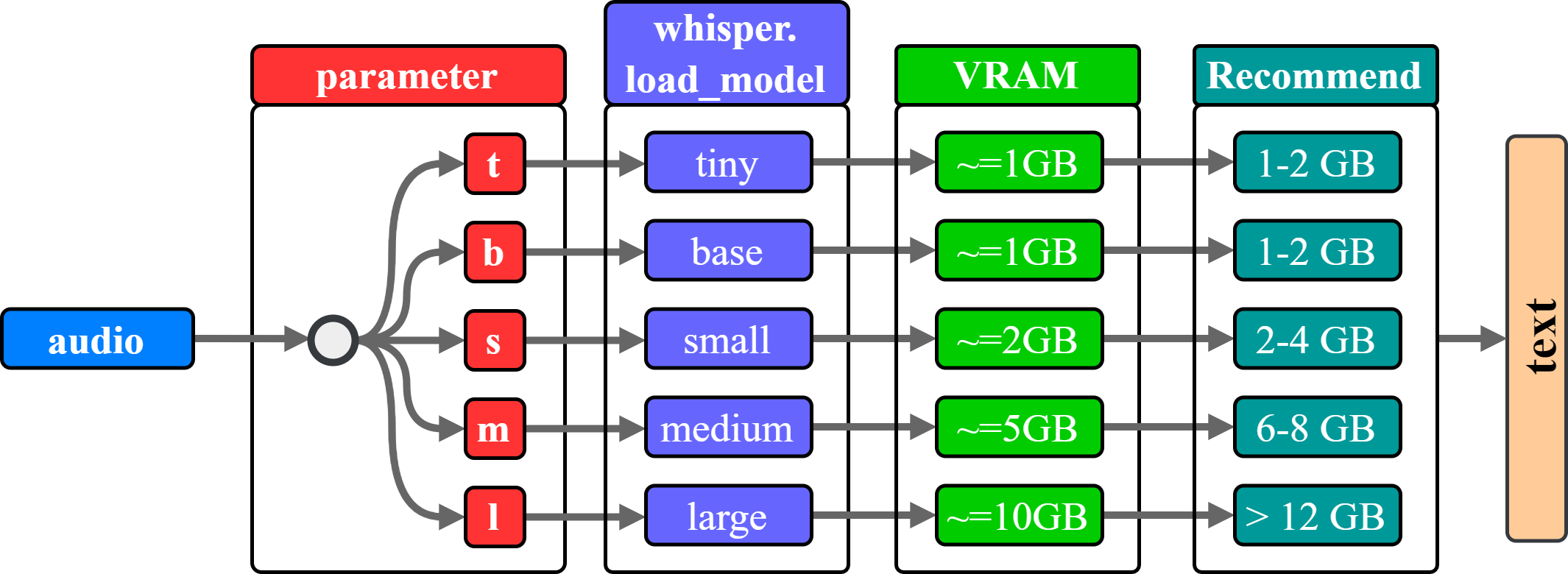}
	\caption{Using the Whisper to get translate the speech of audio to text.}
	\label{fig4}
\end{figure}

Subsequently, we used whisper to implement the speech recognition. We design a custom selection function with five model parameters: $t, b, s, m, l$ to select the five types of model of whisper by using the $load\_model$ function. Each model from whisper needs a different size of Visual RAM (VRAM) to run. As shown in Figure \ref{fig4}. After a series of experiments, we found that the minimum practical required VRAM is bigger than the requirement provided by whisper, so we provide the recommended size parameters for this, as shown in Figure \ref{fig4}. The output result of this process is the text recognized from the speech audio file, which will be used in the next procedure.

% <<:-----------------------------------------

% >>:-----------------------------------------
\subsection{Redundancy Removal}
\label{cm5}
\begin{figure}[t]
	\centering
	\includegraphics[width=1.0\columnwidth]{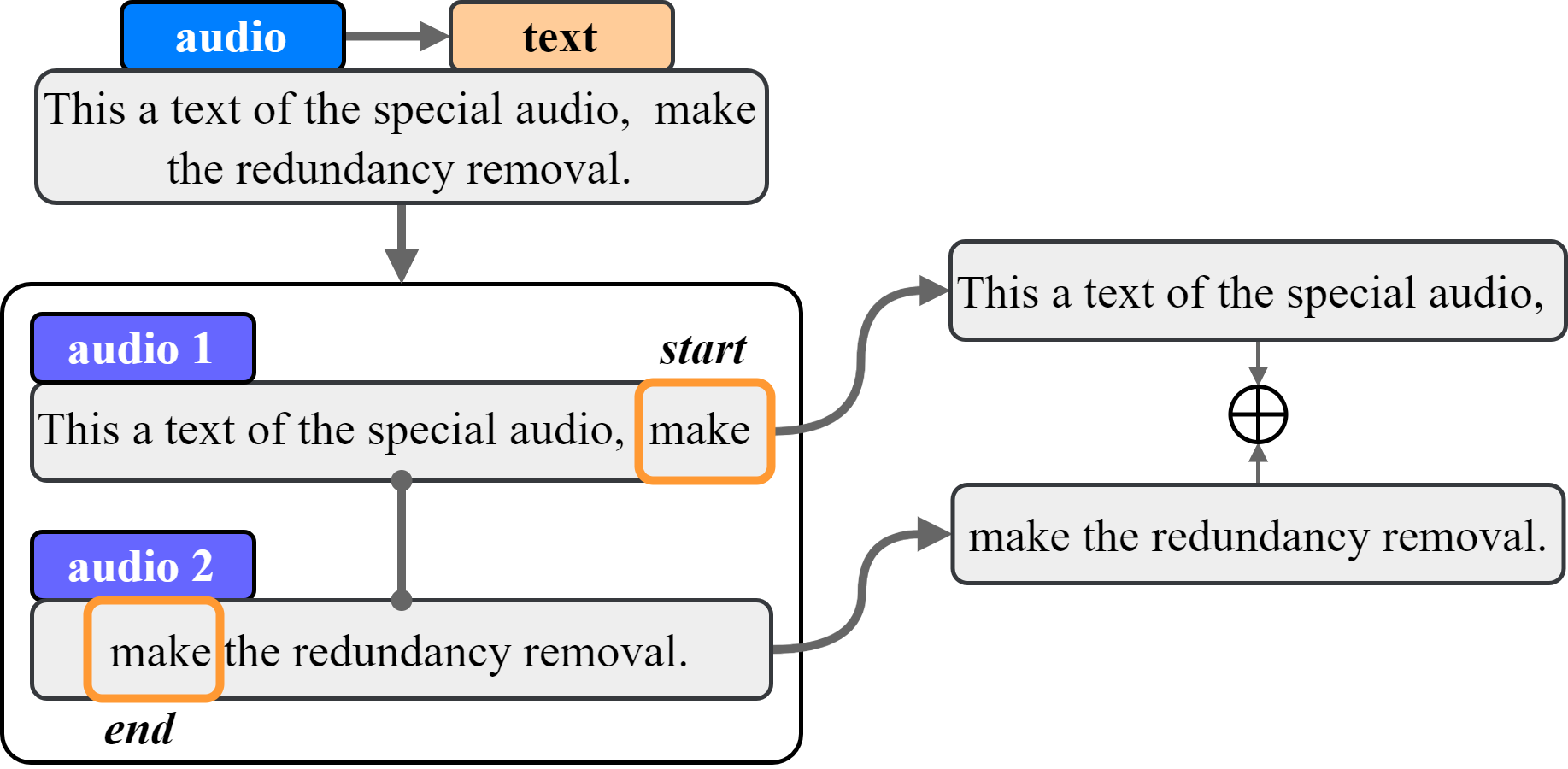}
	\caption{Comparing two closed sentences to make redundancy removal.}
	\label{fig5}
\end{figure}

After getting the text that is recognized and translated from the speech audio, the next procedure is optimising the translated text. In our research, we prepared several methods to make the optimization, of which the first one is, as shown in Figure \ref{fig5}, the redundancy removal of the translated text. The key point of this method is that, comparing the start and end parts of two closed sentences to find the redundant words in some combined context. Because, before the speech audio is sent to make the speech recognition to get the translated text, the speech audio is gained by segment or splitting, which means some repetitive words may be represented in the start and end parts of two closed sentences. Therefore this process will be meaningful to get more accurate translated results in the further speech text processing.

As shown in Figure \ref{fig5}, for example, suppose there is an audio with a speech text “This is a text of the special audio, make the redundancy removal.”, when this audio is recorded or split from the previous process, the audio may be split or recorded into two sentences which have the same repetitive words in the start and end parts of two closed sentences, like $audio 1$ and $audio 2$. Our method is to remain one side of repetitive words in one sentence, and remove the repetitive words in the other sentence. Subsequently, making the further text processing for them, before combining them to get the terminal output. 

% <<:-----------------------------------------

% >>:-----------------------------------------
\subsection{Data Format Conversion}
\label{cm6}
\begin{figure}[t]
	\centering
	\includegraphics[width=1.0\columnwidth]{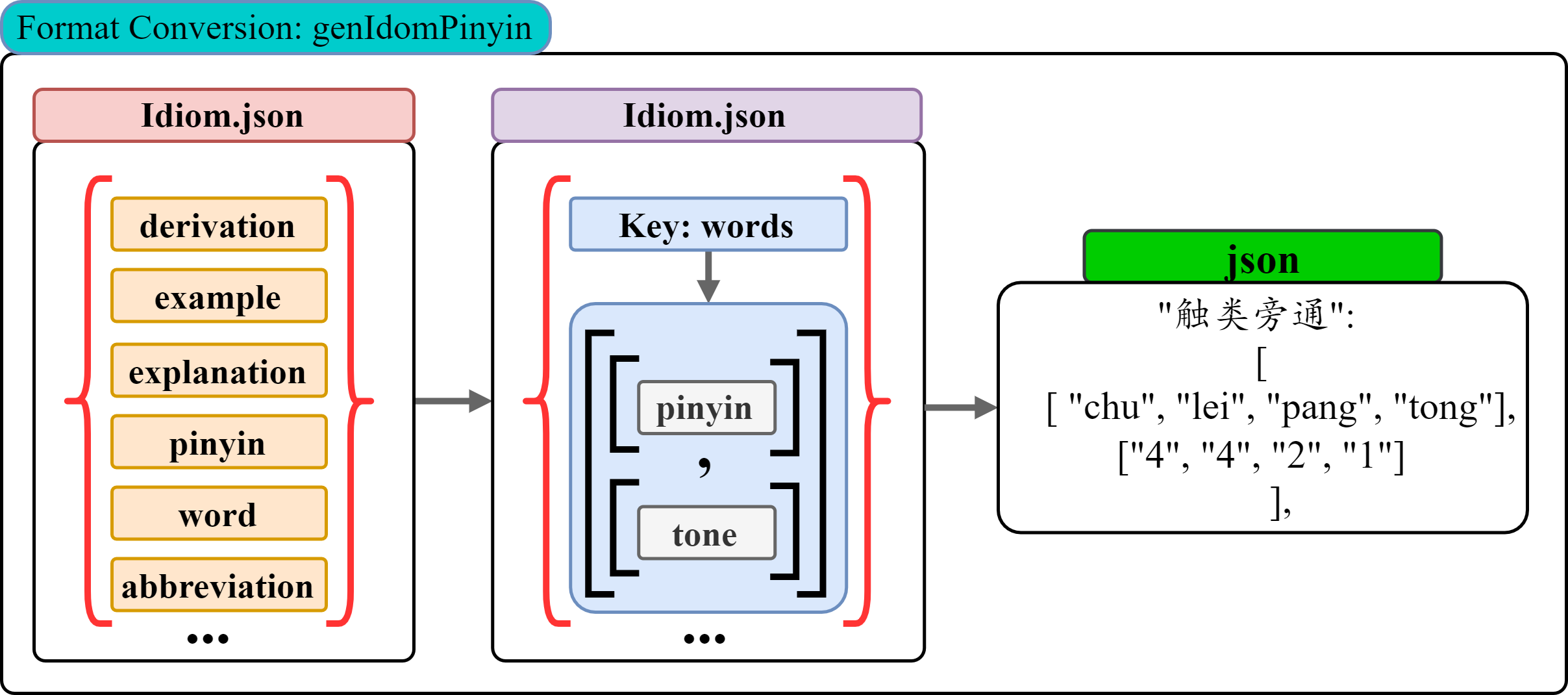}
	\caption{Data format conversion of the idioms to make the dataset.}
	\label{fig6}
\end{figure}

After implementing the redundancy removal of the translated text of the speech audio, the next process for the audio text is building a dataset that follows the special data format with the existing idiom data files. As shown in Figure \ref{fig6}, the previous data format of the idioms json files is combined with:$derivation, example, explanation, pinyin, word$ and $ abbreviation$. Obviously some of these items are non-useful for us to build the idioms for syntax correction. Therefore, we just selected the item$word$ as one component of our new designed idioms dataset. Beside this, the tone of the idiom word is the other required component, which is gained from the program coded by the library $pypinyin$ of Python.

As shown in Figure \ref{fig6}, we design a data format of key-value to store the information of the idioms and their corresponding tones. One unit of this dataset is combined with a Chinese idiom as the kay and a list of  tone information of this idiom as the value. The data format can be described as follows:

\begin{equation}
\label{eq1}
word: \left \{
\begin{matrix} 
[[pinyin_1, pinyin_2, ..., pinyin_n]
 \\ [tone_1, tone_2, ..., tone_n]]]
\end{matrix}, 
\right \} 
\end{equation}

The Equation \ref{eq1} can be used in the data format conversion of other languages that are similar with Chinese, and has phonetic symbols and corresponding tones. With this kind of data format, we can quickly and individually deal with the calculation and comparison between different idiom words, which will promote the further development and research.

% <<:-----------------------------------------

% >>:-----------------------------------------
\subsection{Similarity Computation}
After finishing the building of the dataset of idioms with the  special defined format, the next process is analysing and estimating whether the idiom in the text has a correct expression or not. To address this problem, we design a comparison calculation algorithm to judge and analyse the expression of the idiom. If the expression of the idiom in the text is wrong, the algorithm will find the correct expression from the idiom dataset and replace it in the text with the right expression. 

\begin{figure}[t]
	\centering
	\includegraphics[width=1.0\columnwidth]{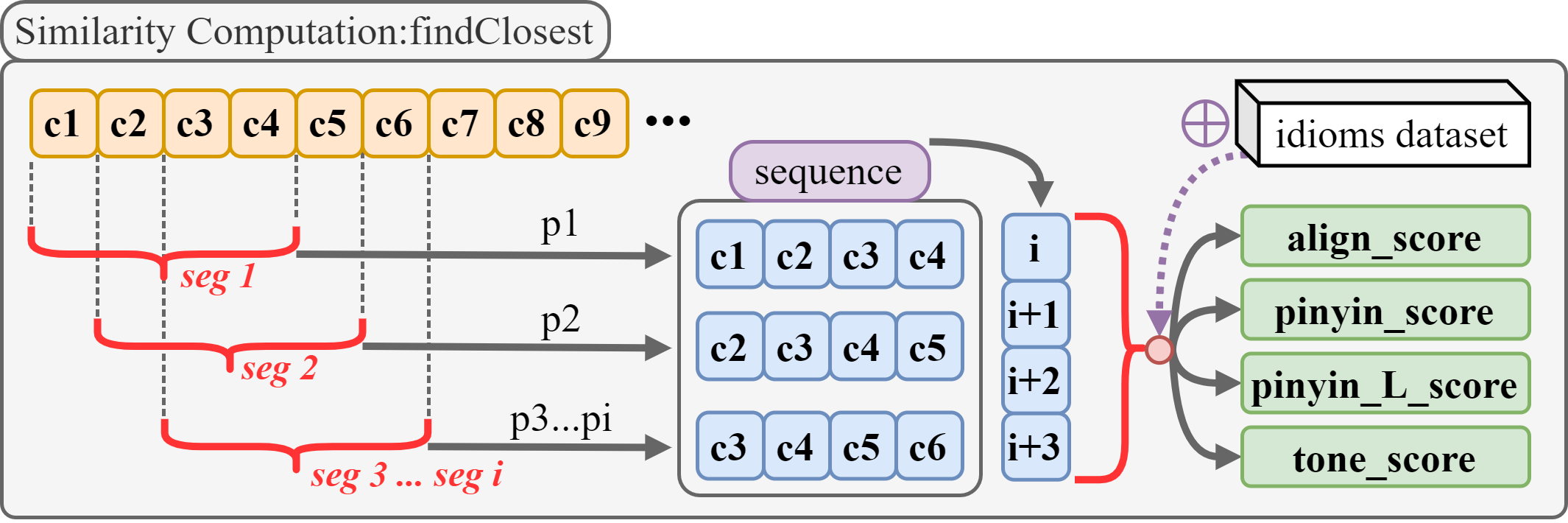}
	\caption{Similarity computation and replacement processing for idioms.}
	\label{fig7}
\end{figure}

To approach this problem, there are four procedures to do. First, get the wrong idiom expression in the audio text. This process is difficult with uncertain interfering factors, for those kinds of languages like Chinese, the words are independent characters, which means it will be difficult to distinguish whether a word character is relative to the word before it or the word behind it. Therefore, a word character in a Chinese sentence, maybe can become an idiom when combining with the words before and after. For this issue, we design an algorithm to analyse the word in a Chinese sentence one by one. As shown in Figure \ref{fig7}, for general Chinese sentences, the character length of a classical Chinese idiom is 4, so with gradual bit positional processing, selecting the sentence segments $seg_1, seg_2, seg_3,  …, seg_i$ to make the positional processing $p_1, p_2, p_3,  …, p_i$ correspondingly. If the current process index of the sentence is $i$, then capture a segment of sentence with a length of 4, which will select the indexes:$i, i+1, i+2, i+3$ to combine a selected sequence $[c_i, c_{i+1}, c_{i+2}, c_{i+3}]$. The length of the idiom is set to be 4, and can also be other numbers like $n, n \in \mathbb{N}^+$.

As shown in the following code, the $word_1$ represents the sequence extracted from the sentence and $word_2$ represents the idiom selected from idioms dataset. $max\_L1$ means the length of the sequence is bigger than the idiom. The judge criteria is relative to length of $word_2$ and parameter $align$. 

\begin{python}
if len(words1) >= len(words2):
    max_L1 = True
else:
    max_L1 = False
    
if max_L1:
    for word_i in range(len(words2)):
        if words1[word_i] == words2[word_i]:
            align_score += a

        if align_score < int(len(words2) / align):
            return_ok = False
else:
    for word_i in range(len(words1)):
        if words1[word_i] == words2[word_i]:
            align_score += a

    if align_score < int(len(words1) / align):
        return_ok = False 
\end{python}

When an idiom expression with a length of $n (n \in \mathbb{N}^+)$ is gained, the next process is calculating the similarity score $S_s$ of the idiom expression with the idiom dataset, to test whether the idiom expression is correct or wrong. We design four kinds of sub-similarity-scores to calculate the total similarity score $S_a$. As shown in Figure \ref{fig7}, the four scores are: $align\_score, pinyin\_socre, pinyin\_L\_score$ and the $tone\_score$. 

\begin{equation}
\label{eq2}
 {\textstyle S_a = \sum_{i}^{n}} (\left \{
\begin{matrix} if \ c_{si} = c_{di}:a
 \\if \ c_{si} \neq c_{di}:0
\end{matrix}       
\right \} )
\end{equation}

In Equation \ref{eq2}, there is the following relationship, the parameter $a=1, (a \in \mathbb{N}^+)$ is the proportional control value,  $c_{si}\in[c_{s1}, c_{s2}, ..., c_{sn}], c_{di}\in[c_{d1}, c_{d2}, ..., c_{dn}$, the parameter $c_{si}$ represents the character on the index $i$ in the $sequence$ selected from the processing sentence. The parameter $c_{di}$ represents the character on the index $i$ in the $idiom$ selected from the idiom dataset. This operation likes bit and operation, if parameter $c_{si}$ equals parameter $c_{di}$, then the total $align\_score$ which represented as $S_s$, will add $a=1$, else will add $0$, that means no more transformation.

Second, for each sequence selected from the sentence, we need to calculate the second evaluation score, that is $pinyin\_score$, which will follow the operation idea as the $align\_score$ and calculate the similarity of the pinyin character by bit. The calculation operation is to compare and calculate the similarity between the pinyin expression of the sequence selected from the sentence and the idiom selected from the idioms dataset. Calculating the similarity between two item that all have the same expression format method in pinyin will be useful for the comparison and similarity calculation of the text correct expression, because the pinyin, as the main articulatory representation item of these languages like Chinese, is most relative expression about the pronunciation of human speech recognition.  Therefore, the parameter $pinyin\_score$ is useful for our similarity replacement algorithm.

\begin{equation}
\label{eq3}
 {\textstyle S_p =  \sum_{i}^{n}} (\left \{
\begin{matrix} if \ cp_{si} = cp_{di}:b
 \\if \ cp_{si} \neq cp_{di}:0
\end{matrix}       
\right \} )
\end{equation}

In Equation \ref{eq3}, there is the following relationship, the parameter $b=1, (b \in \mathbb{N}^+)$ is the proportional control value,  $cp_{si}\in[cp_{s1}, cp_{s2}, ..., cp_{sn}], cp_{di}\in[cp_{d1}, cp_{d2}, ..., cp_{dn}$, the parameter $cp_{si}$ represents the pinyin character on the index $i$ in the pinyin expression of the $sequence$ selected from the processing sentence. The parameter $cp_{di}$ represents the pinyin character on the index $i$ in the pinyin expression of  $idiom$ selected from the idiom dataset. This operation likes bit and operation, if parameter $cp_{si}$ equals parameter $cp_{di}$, then the total $pinyin\_score$ which represented as $S_p$, will add $b=1$, else will add $0$, that means no more transformation. 

Third, besides the small-grained comparison and calculation  mentioned above, we design another calculation algorithm for the large  comparison and for the pinyin expression of each sequence and idiom. For each sequence selected from the sentence, we need to calculate the third evaluation score, that is $pinyin\_L\_score$, which will calculate the similarity of the pinyin character in the unit level of an individual word of the expression of sequence or idiom . The calculation operation is to compare and calculate the similarity between the word pinyin expression of the sequence selected from the sentence and the word pinyin expression of idiom selected from the idioms dataset. Calculating the similarity between two word items to test whether these two expressions are equal will be useful for the comparison and similarity calculation of the text correct expression, because the comparison for the word pinyin expressions between these two items will quickly get the similarity level in the big level of word. The $L$ in $pinyin\_L\_score$ means the $large level$, this is the word level. 

\begin{equation}
\label{eq4}
 {\textstyle S_pl =  \sum_{i}^{n}} (\left \{
\begin{matrix} if \ cpl_{si} = cpl_{di}:c
 \\if \ cpl_{si} \neq cpl_{di}:-c
\end{matrix}       
\right \} )
\end{equation}

In Equation \ref{eq4}, there is the following relationship, the parameter $c=2, (c \in \mathbb{N}^+)$ is the proportional control value,  $cpl_{si}\in[cpl_{s1}, cpl_{s2}, ..., cpl_{sn}], cpl_{di}\in[cpl_{d1}, cpl_{d2}, ..., cpl_{dn}$, the parameter $clp_{si}$ represents, in word level, the pinyin unit expression on the index $i$ in the expression of the $sequence$ selected from the processing sentence. The parameter $cpl_{di}$ represents, in word level, the pinyin unit expression on the index $i$ in the pinyin expression of $idiom$ selected from the idiom dataset. If parameter $cpl_{si}$ equals parameter $cpl_{di}$, then the total $pinyin\_L\_score$ which represented as $S_pl$, will add $c=2$, else will add $-c=-2$, that means reverse transformation. 

Last, after the calculation of the part of pinyin, for the other important part of the tone of idiom, we also design an algorithm to calculate the similarity between tone expressions of each sequence and idiom. For each sequence selected from the sentence, we need to calculate the fourth evaluation score, that is $tone\_score$, which will calculate the similarity of the pinyin tone level number of the expressions of sequence or idiom. The calculation operation is to compare and calculate the similarity distance between the word pinyin tone level number of the sequence selected from the sentence and the pinyin tone level number of idiom selected from the idioms dataset. Calculating the similarity distance between two word items to test whether these two pinyin tone level number of items are equal or their distance will be useful for the comparison and similarity calculation of the correct pronunciation of the idiom item. A correct pronunciation is an important issue to ensure the right expression of the idioms.

\begin{equation}
\label{eq5}
S_t = p\cdot  \frac{ {\textstyle \sum_{i}^{n}} (int(tone_{si})-int(tone_{di}))^2)}{n} 
\end{equation}

In Equation \ref{eq5}, there is the following relationship, the parameter $p=1, (p \in \mathbb{N}^+)$ is the proportional control value,  $c_{si}\in[c_{s1}, c_{s2}, ..., c_{sn}], c_{di}\in[c_{d1}, c_{d2}, ...,  cl_{dn}$, the parameter $c_{si}$ represents, the pinyin tone level number of the word in the index$i$ of expression of the $sequence$ extracted from the processing sentence. The parameter $c_{di}$ represents, in word level, the pinyin tone level number of the word in the index$i$ of expression of $idiom$ selected from the idiom dataset. We use the calculation method of square deviation to calculate the offset distance of similarity of pinyin tone of the sequence and idiom. The parameter $p$ is used to fine-tune the effort of the calculator of the similarity distance of pinyin tone.

During the design and implementation process of similarity calculation, double for loop is usually used to make the double ergodic calculation. To ensure the correct calculation, the iteration number should be the smaller level compared with the second for loop. So $min(iteration\ number)$ should be dimension number of first for loop, the other $bigger(iteration\ number)$ should be dimension number of second for loop.

After all calculations of these four scores, the final evaluation similarity score $S$ can be gained as follow as shown in Equation \ref{eq6}:

\begin{equation}
\label{eq6}
\begin{split}
S = pinyin\_score \times  pp - tone\_score \times pt \\
+ pinyin\_L\_score \times  ppl
\end{split}
\end{equation}

Using the $S$ parameter of Equation \ref{eq6} to evaluate the similarity degree between the $sequence$ extracted from the processing sentence and the $idiom$ selected the idioms data set. Each total calculator score $S$ of each similarity calculation all will be saved in a list. Finally, make the operation $max(scores\_list)$ to get the maximum of the total score. Subsequently, make a judgement of the parameter $S$ to decide whether to replace the sequence with the right idiom expression. If the total similarity score is too small, then remain the previous expression of the extracted sequence, otherwise, replace the sequence with the right expression of the idiom selected from idioms data set.
% <<:-----------------------------------------
% <<:===========================================

% >>:===========================================
\section{Experiment and Evaluation}
To evaluate the performance of our process algorithm, we design the following evaluation method to test the performance of $ecVoice$ from different dimensions. 

% >>:-----------------------------------------
\subsection{Idiom Correction by Word Expression}
First we make the detection experiment for the single idiom correction by using almost some pinyin tone and pronunciation but different expressions in word writing. As shown in Table \ref{tb1}, the parameter $word $ represents the writing of the sequence extracted from the sentence. Due to the sequence word maybe having different expressions in typing or characters writing, we use the pinyin characters and the $\_$ symbol to connect different pinyin of different words. The parameter $n, n \in \mathbb{N}^+$. represents $group\_n$, which means the number of the items of a test group of sequence words. All of the items in the group of sequence words will be tested with our algorithm and count the number of correct test samples, of which number is recorded with the parameter $c, c \in \mathbb{N}^+$.  The parameter $time, time \in \mathbb{N}$ represents the time in seconds of the whole implementation of the group of sequence words. The parameter $s1, s1 \in \mathbb{N}$ represents the similarity score between the correct sequence and the correct idiom, which means the score $s1$ should be the largest one among the similarity scores of the sequence group. Because the first item in the sequence group has the same typing or writing expression of the idiom from the idioms dataset. The parameters $s2, s3 (s2, s3 \in \mathbb{N})$ represent the similarity scores between two incorrect sequences and the correct idiom from the idioms dataset. Therefore parameters $s2$ and $s3$ should be never larger than $s1$. The parameter $sa, sa \in \mathbb{N}$ represents the average of the parameters $s2$ and $s3$, which represents the average of the normal similarity score of this sequence test group.

\begin{table}[htb]   
\begin{center}   
\caption{word error index = 1.}  
\label{tb1} 
\begin{tabular}{|c|c|c|c|c|c|c|c|c|}   
\hline   \textbf{word} & \textbf{n} & \textbf{c} & \textbf{time} & \textbf{s1} & \textbf{s2} & \textbf{s3} & \textbf{sa} \\   

\hline ri\_chu\_san\_gan         & 3 & 3 & 1.207 & 19.0 & 19.0 & 14.0 & 16.5    \\
\hline chu\_mu\_cheng\_song      & 3 & 3 & 1.18 & 22.0 & 16.0 & 22.0 & 19.0     \\
\hline fang\_ge\_zong\_jiu       & 3 & 3 & 1.156 & 21.0 & 21.0 & 18.75 & 19.9   \\
\hline jing\_bing\_qiang\_jiang  & 3 & 3 & 1.203 & 26.0 & 23.75 & 23.75 & 23.8  \\
\hline man\_shan\_bian\_ye       & 3 & 3 & 1.151 & 21.0 & 21.0 & 21.0 & 21.0    \\
\hline po\_bu\_ji\_dai           & 3 & 3 & 1.123 & 17.0 & 17.0 & 17.0 & 17.0    \\
\hline qing\_chu\_yu\_lan        & 3 & 3 & 1.121 & 20.0 & 20.0 & 20.0 & 20.0    \\
\hline ren\_shou\_nian\_feng     & 3 & 3 & 1.135 & 23.0 & 23.0 & 23.0 & 23.0    \\
\hline shi\_wai\_tao\_yuan       & 3 & 3 & 1.127 & 21.0 & 21.0 & 21.0 & 21.0    \\
\hline wu\_ti\_tou\_di           & 3 & 3 & 1.126 & 17.0 & 17.0 & 17.0 & 17.0    \\

\hline   
\end{tabular}   
\end{center}   
\end{table}

As shown in Table \ref{tb1}, the word error index is 1, that means for all sequences with the incorrect expression of the test sequence group, the first character is wrong for the idiom expression. Here for convenience to represent the word site of the sequence, the number of the index starts from 1. The result shows all of the sequences with the wrong idiom expression have been detected. For different sequences, the similarity score used to judge is different. The approximate similarity score is in the range of from 17.0 to 26.0. The implementation time for each sequence group is about 1.2 seconds.

Subsequently, we implement the experiment of sequence groups that have the wrong textual expression sequences in the index of 2, which means the second characters of those sequences are in the wrong expression. As shown in Table \ref{tb2}, the result shows all of the sequences with the wrong idiom expression have been detected. The approximate similarity score is in the range of from 17.0 to 26.0. The implementation time for each sequence group is about 1.15 seconds.

\begin{table}[htb]   
\begin{center}   
\caption{word error index = 2.}  
\label{tb2} 
\begin{tabular}{|c|c|c|c|c|c|c|c|c|}   
\hline   \textbf{word} & \textbf{n} & \textbf{c} & \textbf{time} & \textbf{s1} & \textbf{s2} & \textbf{s3} & \textbf{sa} \\   

\hline ri\_chu\_san\_gan            & 3 & 3 & 1.149 & 19.0 & 19.0 & 19.0 & 19.0  \\
\hline chu\_mu\_cheng\_song         & 3 & 3 & 1.166 & 22.0 & 22.0 & 22.0 & 22.0  \\
\hline fang\_ge\_zong\_jiu          & 3 & 3 & 1.255 & 21.0 & 18.75 & 18.75 & 18.8  \\
\hline jing\_bing\_qiang\_jiang     & 3 & 3 & 1.27 & 26.0 & 23.75 & 23.75 & 23.8  \\
\hline man\_shan\_bian\_ye          & 3 & 3 & 1.209 & 21.0 & 21.0 & 21.0 & 21.0  \\
\hline po\_bu\_ji\_dai              & 3 & 3 & 1.192 & 17.0 & 17.0 & 17.0 & 17.0  \\
\hline qing\_chu\_yu\_lan           & 3 & 3 & 1.147 & 20.0 & 20.0 & 20.0 & 20.0  \\
\hline ren\_shou\_nian\_feng        & 3 & 3 & 1.156 & 23.0 & 23.0 & 23.0 & 23.0  \\
\hline shi\_wai\_tao\_yuan          & 3 & 3 & 1.137 & 21.0 & 21.0 & 21.0 & 21.0  \\
\hline wu\_ti\_tou\_di              & 3 & 3 & 1.089 & 17.0 & 17.0 & 17.0 & 17.0  \\

\hline   
\end{tabular}   
\end{center}   
\end{table}

Next, we implement the experiment of sequence groups that have the wrong textual expression sequences in the index of 3, which means the third characters of those sequences are in the wrong expression. As shown in Table \ref{tb3}, the result shows all of the sequences with the wrong idiom expression have been detected. The approximate similarity score is in the range of from 17.0 to 26.0. The implementation time for each sequence group is about 1.15 seconds.

\begin{table}[htb]   
\begin{center}   
\caption{word error index = 3.}  
\label{tb3} 
\begin{tabular}{|c|c|c|c|c|c|c|c|c|}   
\hline   \textbf{word} & \textbf{n} & \textbf{c} & \textbf{time} & \textbf{s1} & \textbf{s2} & \textbf{s3} & \textbf{sa} \\   

\hline ri\_chu\_san\_gan           & 3 & 3 & 1.149 & 19.0 & 19.0 & 19.0 & 19.0  \\
\hline chu\_mu\_cheng\_song        & 3 & 3 & 1.166 & 22.0 & 22.0 & 22.0 & 22.0  \\
\hline fang\_ge\_zong\_jiu         & 3 & 3 & 1.255 & 21.0 & 18.75 & 18.75 & 18.8  \\
\hline jing\_bing\_qiang\_jiang    & 3 & 3 & 1.27 & 26.0 & 23.75 & 23.75 & 23.8  \\
\hline man\_shan\_bian\_ye         & 3 & 3 & 1.209 & 21.0 & 21.0 & 21.0 & 21.0  \\
\hline po\_bu\_ji\_dai             & 3 & 3 & 1.192 & 17.0 & 17.0 & 17.0 & 17.0  \\
\hline qing\_chu\_yu\_lan          & 3 & 3 & 1.147 & 20.0 & 20.0 & 20.0 & 20.0  \\
\hline ren\_shou\_nian\_feng       & 3 & 3 & 1.156 & 23.0 & 23.0 & 23.0 & 23.0  \\
\hline shi\_wai\_tao\_yuan         & 3 & 3 & 1.137 & 21.0 & 21.0 & 21.0 & 21.0  \\
\hline wu\_ti\_tou\_di             & 3 & 3 & 1.089 & 17.0 & 17.0 & 17.0 & 17.0  \\

\hline   
\end{tabular}   
\end{center}   
\end{table}

Subsequently, we implement the experiment of sequence groups that have the wrong textual expression sequences in the index of 4, which means the fourth characters of those sequences are in the wrong expression. As shown in Table \ref{tb4}, the result shows all of the sequences with the wrong idiom expression have been detected. The approximate similarity score is in the range of from 17.0 to 26.0. The implementation time for each sequence group is about 1.14 seconds.

\begin{table}[htb]   
\begin{center}   
\caption{word error index = 4.}  
\label{tb4} 
\begin{tabular}{|c|c|c|c|c|c|c|c|c|}   
\hline   \textbf{word} & \textbf{n} & \textbf{c} & \textbf{time} & \textbf{s1} & \textbf{s2} & \textbf{s3} & \textbf{sa} \\   

\hline ri\_chu\_san\_gan            & 3 & 3 & 1.159 & 19.0 & 19.0 & 19.0 & 19.0   \\
\hline chu\_mu\_cheng\_song         & 3 & 3 & 1.149 & 22.0 & 22.0 & 22.0 & 22.0   \\
\hline fang\_ge\_zong\_jiu          & 3 & 3 & 1.155 & 21.0 & 21.0 & 21.0 & 21.0   \\
\hline jing\_bing\_qiang\_jiang     & 3 & 3 & 1.259 & 26.0 & 23.75 & 23.75 & 23.8   \\
\hline man\_shan\_bian\_ye          & 3 & 3 & 1.152 & 21.0 & 21.0 & 21.0 & 21.0   \\
\hline po\_bu\_ji\_dai              & 3 & 3 & 1.137 & 17.0 & 17.0 & 17.0 & 17.0   \\
\hline qing\_chu\_yu\_lan           & 3 & 3 & 1.107 & 20.0 & 20.0 & 20.0 & 20.0   \\
\hline ren\_shou\_nian\_feng        & 3 & 3 & 1.146 & 23.0 & 23.0 & 23.0 & 23.0   \\
\hline shi\_wai\_tao\_yuan          & 3 & 3 & 1.152 & 21.0 & 21.0 & 21.0 & 21.0   \\
\hline wu\_ti\_tou\_di              & 3 & 3 & 1.096 & 17.0 & 17.0 & 17.0 & 17.0   \\

\hline   
\end{tabular}   
\end{center}   
\end{table}

The operations mentioned above is all about controlling one of the characters of the sequence to be wrong in expression, followed by detecting and calculating the similarity between the sequence with wrong expression and the idiom with correct expression. To better test the performance of our algorithm, we design the following experiments, which set   two characters of the sequence to be wrong and test the similarity score. As shown in Table \ref{tb5}, setting the first and second characters of the sequences to be wrong and then testing the similarity score, meanwhile recording the implementation time for each sequence group. As shown in Table \ref{tb5}, the result shows all of the sequences with the wrong idiom expression have been detected. The approximate similarity score is in the range of from 17.0 to 26.0. The implementation time for each sequence group is about 1.14 seconds.

\begin{table}[htb]   
\begin{center}   
\caption{word error index = 1,2.}  
\label{tb5} 
\begin{tabular}{|c|c|c|c|c|c|c|c|c|}   
\hline   \textbf{word} & \textbf{n} & \textbf{c} & \textbf{time} & \textbf{s1} & \textbf{s2} & \textbf{s3} & \textbf{sa} \\   

\hline ri\_chu\_san\_gan             & 3 & 3 & 1.161 & 19.0 & 19.0 & 14.0 & 16.5  \\
\hline chu\_mu\_cheng\_song          & 3 & 3 & 1.188 & 22.0 & 16.0 & 22.0 & 19.0  \\
\hline fang\_ge\_zong\_jiu          & 3 & 3 & 1.142 & 21.0 & 21.0 & 18.75 & 19.9  \\
\hline jing\_bing\_qiang\_jiang      & 3 & 3 & 1.155 & 26.0 & 26.0 & 21.0 & 23.5  \\
\hline man\_shan\_bian\_ye           & 3 & 3 & 1.168 & 21.0 & 21.0 & 21.0 & 21.0  \\
\hline po\_bu\_ji\_dai               & 3 & 3 & 1.133 & 17.0 & 17.0 & 17.0 & 17.0  \\
\hline qing\_chu\_yu\_lan           & 3 & 3 & 1.126 & 20.0 & 20.0 & 19.75 & 19.9  \\
\hline ren\_shou\_nian\_feng          & 3 & 3 & 1.13 & 23.0 & 23.0 & 23.0 & 23.0  \\
\hline shi\_wai\_tao\_yuan           & 3 & 3 & 1.143 & 21.0 & 21.0 & 21.0 & 21.0  \\
\hline wu\_ti\_tou\_di               & 3 & 3 & 1.086 & 17.0 & 17.0 & 17.0 & 17.0  \\

\hline   
\end{tabular}   
\end{center}   
\end{table}

After testing the influence of the two characters in the start of the sequence to the similarity score calculation, we have also tested the influence of the two characters in the end of the sequence to the similarity score calculation. As shown in Table \ref{tb6}, setting the third and fourth characters of the sequences to be wrong and then testing the similarity score, meanwhile recording the implementation time for each sequence group. As shown in Table \ref{tb6}, the result shows all of the sequences with the wrong idiom expression have been detected. The approximate similarity score is in the range of from 17.0 to 26.0. The implementation time for each sequence group is about 1.25 seconds.

\begin{table}[htb]   
\begin{center}   
\caption{word error index = 3,4.}  
\label{tb6} 
\begin{tabular}{|c|c|c|c|c|c|c|c|c|}   
\hline   \textbf{word} & \textbf{n} & \textbf{c} & \textbf{time} & \textbf{s1} & \textbf{s2} & \textbf{s3} & \textbf{sa} \\   

\hline ri\_chu\_san\_gan            & 3 & 3 & 1.163 & 19.0 & 19.0 & 19.0 & 19.0   \\
\hline chu\_mu\_cheng\_song         & 3 & 3 & 1.154 & 22.0 & 22.0 & 22.0 & 22.0   \\
\hline fang\_ge\_zong\_jiu        & 3 & 3 & 1.222 & 21.0 & 18.75 & 18.75 & 18.8   \\
\hline jing\_bing\_qiang\_jiang    & 3 & 3 & 1.24 & 26.0 & 23.75 & 23.75 & 23.8   \\
\hline man\_shan\_bian\_ye          & 3 & 3 & 1.646 & 21.0 & 21.0 & 21.0 & 21.0   \\
\hline po\_bu\_ji\_dai               & 3 & 3 & 1.14 & 17.0 & 17.0 & 17.0 & 17.0   \\
\hline qing\_chu\_yu\_lan           & 3 & 3 & 1.111 & 20.0 & 20.0 & 20.0 & 20.0   \\
\hline ren\_shou\_nian\_feng        & 3 & 3 & 1.141 & 23.0 & 23.0 & 23.0 & 23.0   \\
\hline shi\_wai\_tao\_yuan          & 3 & 3 & 1.185 & 21.0 & 21.0 & 21.0 & 21.0   \\
\hline wu\_ti\_tou\_di              & 3 & 3 & 1.148 & 17.0 & 17.0 & 17.0 & 17.0   \\

\hline   
\end{tabular}   
\end{center}   
\end{table}
% <<:-----------------------------------------

% >>:-----------------------------------------
\subsection{Idiom Correction by Tone Pronunciation}
Due the pinyin tone of the sequence is another important factor for the similarity calculation for our similarity analysis and replacement algorithm, we implemented the following two kinds of experiments to test the performance of our algorithm under the influence of wrong pinyin tone. Because the tone of pinyin can make the word character, so in this condition, the textual expression of the sequence usually are different.  

As shown in Table \ref{tb7}, setting the first and second tone numbers of the sequence pinyin to be wrong and then testing the similarity score, meanwhile recording the implementation time for each sequence group. As shown in Table \ref{tb7}, the result shows all of the sequences with the wrong idiom expression have been detected. The approximate similarity score is in the range of from 8.75 to 21.75. The implementation time for each sequence group is about 1.30 seconds.

\begin{table}[htb]   
\begin{center}   
\caption{tone error index = 1,2.}  
\label{tb7} 
\begin{tabular}{|c|c|c|c|c|c|c|c|c|}   
\hline   \textbf{word} & \textbf{n} & \textbf{c} & \textbf{time} & \textbf{s1} & \textbf{s2} & \textbf{s3} & \textbf{sa} \\   

\hline ri\_chu\_san\_gan           & 3 & 3 & 1.169 & 19.0 & 12.75 & 11.5 & 12.1   \\
\hline chu\_mu\_cheng\_song         & 3 & 3 & 1.293 & 22.0 & 8.75 & 10.75 & 9.8   \\
\hline fang\_ge\_zong\_jiu         & 3 & 3 & 1.233 & 21.0 & 11.5 & 15.75 & 13.6   \\
\hline jing\_bing\_qiang\_jiang     & 3 & 3 & 1.18 & 26.0 & 20.0 & 17.75 & 18.9   \\
\hline man\_shan\_bian\_ye         & 3 & 1 & 1.154 & 21.0 & 9.75 & 11.75 & 10.8   \\
\hline po\_bu\_ji\_dai             & 3 & 2 & 1.139 & 17.0 & 16.75 & 5.75 & 11.2   \\
\hline qing\_chu\_yu\_lan           & 3 & 3 & 1.152 & 20.0 & 14.5 & 15.5 & 15.0   \\
\hline ren\_shou\_nian\_feng       & 3 & 3 & 1.144 & 23.0 & 21.75 & 15.0 & 18.4   \\
\hline shi\_wai\_tao\_yuan         & 3 & 3 & 1.114 & 21.0 & 7.75 & 12.75 & 10.2   \\
\hline wu\_ti\_tou\_di              & 3 & 3 & 1.101 & 17.0 & 16.5 & 16.5 & 16.5   \\

\hline   
\end{tabular}   
\end{center}   
\end{table}

As shown in Table \ref{tb8}, setting the third and fourth tone numbers of the sequence pinyin to be wrong and then testing the similarity score, meanwhile recording the implementation time for each sequence group. As shown in Table \ref{tb8}, the result shows all of the sequences with the wrong idiom expression have been detected. The approximate similarity score is in the range of from 8.5 to 25.5. The implementation time for each sequence group is about 1.35 seconds.

\begin{table}[htb]   
\begin{center}   
\caption{tone error index = 3,4.}  
\label{tb8} 
\begin{tabular}{|c|c|c|c|c|c|c|c|c|}   
\hline   \textbf{word} & \textbf{n} & \textbf{c} & \textbf{time} & \textbf{s1} & \textbf{s2} & \textbf{s3} & \textbf{sa} \\   

\hline ri\_chu\_san\_gan             & 3 & 3 & 1.146 & 19.0 & 18.0 & 8.5 & 13.2   \\
\hline chu\_mu\_cheng\_song       & 3 & 3 & 1.153 & 22.0 & 14.75 & 21.75 & 18.2   \\
\hline fang\_ge\_zong\_jiu         & 3 & 3 & 1.162 & 21.0 & 10.75 & 18.5 & 14.6   \\
\hline jing\_bing\_qiang\_jiang       & 3 & 3 & 1.2 & 26.0 & 25.5 & 23.5 & 24.5   \\
\hline man\_shan\_bian\_ye         & 3 & 3 & 1.182 & 21.0 & 18.75 & 20.5 & 19.6   \\
\hline po\_bu\_ji\_dai              & 3 & 3 & 1.129 & 17.0 & 16.5 & 14.5 & 15.5   \\
\hline qing\_chu\_yu\_lan          & 3 & 3 & 1.149 & 20.0 & 18.75 & 19.0 & 18.9   \\
\hline ren\_shou\_nian\_feng      & 3 & 3 & 1.141 & 23.0 & 15.75 & 22.75 & 19.2   \\
\hline shi\_wai\_tao\_yuan         & 3 & 3 & 1.147 & 21.0 & 19.75 & 20.5 & 20.1   \\
\hline wu\_ti\_tou\_di             & 3 & 3 & 1.102 & 17.0 & 15.75 & 16.5 & 16.1   \\

\hline   
\end{tabular}   
\end{center}   
\end{table}

% <<:-----------------------------------------

% >>:-----------------------------------------
\subsection{Optimization for Audio Text Recognition}
To test the optimization and assistance of our whole algorithm to the audio text recognition, we implemented the following comparison experiments. We use some video selected by random and use our algorithms to make the audio text recognition, comparison the output of the methods and the original output from the small size of the whisper model.

\begin{table}[htb]   
\begin{center}   
\caption{model or method performances comparison}  
\label{tb9} 
\begin{tabular}{|c|c|c|c|c|c|c|c|c|}   
\hline   \textbf{methods} & \textbf{performance} & \textbf{useful} \\   

\hline HanLP & ~60-85\% & less useful   \\
\hline original whisper & ~81\% & normal  \\
\hline our ecVoice & ~81-95\% & useful  \\

\hline   
\end{tabular}   
\end{center}   
\end{table}

As shown in Table \ref{tb9}, HanLP-v1.8.4 \cite{he-choi-2021-stem} is very famous Chinese syntax correction and NLP process framework, but in special application like idiom syntax problem analysis or detection is needed more update. Our algorithm performances well in this issue, which performance good in small size model.

% <<:-----------------------------------------
% <<:===========================================

% >>:===========================================
\section{Conclusion}
By the experiments analysis mentioned above, we can get this conclusion: the similarity calculator and judgement part of the whole similarity replacement algorithm has a high efficiency. The running time for each sequence group is about 1.5 seconds, which can be optimised by upgrading the search algorithm in the further research. Compared with tone wrong expression detection, the  word wrong expression detection needs less implementation time and has a higher efficiency. For application situations that without  non-real-time requirements, this still can provide a strong and fast process performance.
% <<:===========================================

% >>:====================================================
\bibliographystyle{IEEEtran}
\bibliography{ref}{}
\end{document}
